\documentstyle[prb,aps,epsf,twocolumn]{revtex}
\begin{document}

\hfuzz=100pt
\hbadness=10000

\twocolumn[\hsize\textwidth\columnwidth\hsize\csname @twocolumnfalse\endcsname

\title{\bf Changing shapes in the nanoworld}

\author{Nicolas Combe (a), Pablo Jensen (a) and Alberto Pimpinelli (b)}
\address{(a) D\'epartement de Physique des
Mat\'eriaux, UMR CNRS 5586, Universit\'e Claude Bernard Lyon-1, 69622
Villeurbanne C\'edex, FRANCE;
\\ (b) LASMEA, Universit\'e Blaise Pascal Clermont-2, Les
C\'ezeaux, 63177 Aubi\`ere C\'edex, FRANCE}

\maketitle

\begin{abstract} 
What are the mechanisms leading to the shape relaxation of three
dimensional crystallites ? Kinetic Monte Carlo simulations of fcc
clusters show that the usual theories
of equilibration, via atomic surface diffusion driven by curvature,
are verified only at high temperatures. Below the roughening
temperature, the relaxation is much slower, kinetics being
governed by the nucleation of a critical germ on a facet. We show that 
the energy barrier for this
step linearly increases with the size of the crystallite, leading to
an exponential dependence of the relaxation time.
\end{abstract}

\vskip2pc]

\narrowtext

Imagine a world where marbles would fuse upon contact, just as two
water droplets usually do to minimize their surface energy and reach
their equilibrium configuration. This peculiar behavior is thought to
be usual in the {\it nano}world, because experiments suggest that
objects in the nanometer range can change shape in reasonable times
\cite{exp,review}, even if they are {\it solid}. This fact is
crucial for the production and control of nanostructures, since these are
generally obtained in out-of-equilibrium conditions, and are therefore 
metastable. The rapid shape relaxation of these
solid particles (which lack the collective atomic diffusion mechanisms
found in the liquid state) is dominated by
surface diffusion of their surface atoms. Atomic diffusion is
random (brownian) but nevertheless generates a global mass transfer
from the high curvature regions (of higher chemical potential, roughly
because atoms have less neighbors there) to the low curvature regions. For 
very small objects, surface diffusion is very efficient as a
mass transfer mechanism. However, it has to be pointed out that this
whole picture assumes that a continuous description of the objects is
valid, thus allowing the definition of a curvature-dependent local
chemical potential. This is correct as long as the particle is large 
enough (to allow defining a chemical potential) and has a disordered or 
rough surface (in order that the chemical potential is differentiable) as 
for a liquid droplet. Within these assumptions, Herring, Nichols and 
Mullins \cite{nichols} have shown that this
mass transfer mechanism leads to an equilibrium time $t_{eq}$ which
increases as the fourth power of the object linear size, which helps
understanding why macroscopic objects are kinetically allowed to
violate thermodynamics with total impunity. It is important to note
that this fourth power law is extensively used to predict equilibrium
times or to extract diffusion constants from equilibration rates
\cite{review,expmull}. In this paper, we address the important question of 
what happens when the temperature is below the roughening transition 
\cite{alb}, so that the particle is faceted, a common situation for 
nanostructures. There is no general agreement on a continous approach
in this regime
\cite{rough_cont}, and therefore we use Monte Carlo simulations to show 
that the equilibration time of a faceted particle does not simply vary 
as predicted above. Indeed, we show that, although surface diffusion 
remains the important matter transport channel, the equilibration time 
is limited by another physical mechanism -- facet nucleation. This different 
physical mechanism leads to a very different size dependence of $t_{eq}$,
limiting the validity of the continuous law and explaining why nanobjects
might spend much more time in metastable equilibrium than expected.

\vspace{.4cm}
\fbox{\it Physical model}
\vspace{.4cm}

We use standard Kinetic Monte Carlo simulations 
\cite{kmc} to study the equilibration of unsupported 3D
crystallites having a perfect fcc crystalline structure.  Since we are
only interested in finding generic laws for the size dependence of
$t_{eq}$ (which should not depend on the details of atom-atom interaction), 
we have chosen a very simple energy landscape for atomic
motion \cite{last_one}.  We assume that the potential energy 
$E_p$ of an atom is
proportional to its number $i$ of neighbors, and that the {\it kinetic
barrier} $E_{act}$ for diffusion is also proportional to the number of
{\it initial} neighbors, before the jump, regardless of the {\it
final} number of neighbors, after the jump \cite{note1} : $ E_{act}=-
E_p = i*E $ where $E$ sets the energy scale ($E=0.1$ eV throughout the
paper).  Comparing with recent ab-initio calculations
\cite{bogicevic} for the Al(111) surface, we note that our one-barrier
assumption does give the good order of magnitude of the relative jump
frequencies for the different hopping processes of interest here. We
also exclude any explicit ``Ehrlich-Schwoebel" barrier \cite{es} for
atoms hopping around corners. Therefore, the
probability $p_i$ per unit time that an atom with $i$ neighbors
moves is $p_i = \nu_0 \exp[-i*E/k_bT]$,
where $\nu_0= 10^{13} s^{-1}$ is the Debye frequency. Thus, our simple
kinetic model includes only {\it one parameter}, the ratio $E/k_B T$
where $k_B$ is the Boltzmann constant and $T$ the absolute
temperature. We have changed the temperature from $300 K$ to $800 K$,
and the crystallite number of atoms from $700$ up to $13000$. The
initial configuration of the clusters is elongated (same initial 
aspect ratio \cite{note3}), and we stop the relaxation when the 
crystallites are close to equilibrium, with an aspect ratio of 1.2.

\vspace{.4cm}
\fbox{\it Simulation results}
\vspace{.4cm}

Fig. \ref{teq_N} shows a log-log plot of the relaxation time as a
function of the number of atoms in the crystallite. The continuous law
predicts a slope of $4/3$, which agrees with our simulations only for
the highest temperature, $800 K$. As the temperature decreases, the
slope {\it continuously increases}, reaching much higher values than
this $4/3$. This strong deviation from the continuous law suggests
that an altogether physical mechanism limits the mass transfer at low
temperature. The constant increase of the exponent is a clue that an
exponential dependence of $t_{eq}$ on size and temperature might be
present.

To investigate the different behaviors at high and low temperatures,
we examine (Fig. \ref{morpho}) the different morphologies of the
clusters at $T=700 K$ and $T=300 K$. At high temperatures, many kinks
and steps can be seen, indicating that the continuous approximation
for the curvature might be valid. On the contrary, at low
temperatures the crystallite is fully faceted, with angular points
and edges, making it difficult to define a chemical potential
properly. Moreover, the presence of facets makes it impossible to
transfer atoms from the cluster tips to its central region by simple
atomic diffusion : atoms reaching the facets do not find a trapping
site there and eventually get back to kinks or steps in the tip
regions. This means that the crystallites can be trapped for long
times in these faceted configurations at low temperatures, as can be
clearly seen on Fig. \ref{energy} : the cluster progressive approach
of equilibrium is continuous at high temperature and more steplike at 
low temperature, indicating that at high temperature atoms continuously 
attach to the central regions, whereas
some more discontinuous mechanism operates at low temperatures. A
careful examination of low temperature relaxation pictures suggested that
the transition from one step (i.e. a fully faceted, relatively stable
configuration) to the following (lower) step
demands the nucleation of a germ on a large facet. This germ
grows and eventually forms a new atomic layer, thus bringing the crystallite 
closer to equilibrium (see Fig. \ref{energy}). Therefore, we expect that the
limiting process for matter transfer at low temperatures is the
formation of a critical nucleus, as suggested from
classical nucleation theory \cite{nucleation}. 

This physical picture can be partly quantified using the kinetic
theory of nucleation \cite{nucleation}. The time
needed to build the critical nucleus is given by :

\begin{equation} 
t_{nucl} \ \propto \  exp(\frac{\Delta G^*}{k_BT})
\label{nucl}
\end{equation}

\noindent
where $\Delta G^*$ is the free energy barrier the system has to cross.
We stress that here the situation is more complex than in the
gas-liquid transition, where the atoms forming the incipient liquid
critical cluster all come from the gas phase, which has a fixed
chemical potential. Here, the tip atoms come from different
environments with different energies (see the different colors in 
Fig. \ref{morpho}), making it difficult to calculate an average chemical 
potential. The key point is however to
examine whether $\Delta G^*$ depends on the particle size, thus
creating an exponential contribution to the size dependence of
$t_{eq}$.

We use a classical umbrella sampling technique \cite{umbrella} to
compute the crystallite free energy as a function of the number of
atoms in the nucleating germ. The umbrella technique consists in
adding a bias potential to the hamiltonian of the system to force it
to stay in a configuration of interest, even if it is
unprobable, as is the case here for nucleation of the germ. 
Fig. \ref{deltag} shows that $\Delta G^*$ increases for
larger crystallites \cite{note2}, which implies (Eq. \ref{nucl}) that the
nucleation time (and therefore $t_{eq}$) depends exponentially on the
size of the cluster (provided, of course, that the $\Delta G^*$
increase is not logarithmic, see below).

What are the microscopic mechanisms leading to this $\Delta G^*$ increase 
with crystallite size? The free energy of a nucleating germ is given by 
\cite{nucleation} :

\begin{equation}
\Delta G = 2 \gamma_{line} \sqrt{\pi q} - q \Delta \mu
\label{fit}
\end{equation}

\noindent
where q is the number of atoms in the germ, $\gamma_{line}$ the
line tension of the germ, and $\Delta \mu$ the chemical potential
difference for an atom going from the tip to the facet. We fit the curves 
of Fig. \ref{deltag} by Eq. \ref{fit} which gives $\gamma_{line}$ 
and $\Delta \mu$. We find $\gamma_{line} = 0.129 \pm 0.013 eV/atom$, 
{\it independent} of the crystallite size. This value is, 
 as expected, close to the binding energy. To understand the size
dependence of  $\Delta G^*$ (which comes from $\Delta \mu$), one can, 
as a first approximation, treat the tips in a continuous way : assimilating 
them to half an ellipsoid, we can estimate the tip curvature. This gives a 
rough measure of the kink and step density
on the tips, and therefore of the density of more or less mobile
atoms, which can contribute to mass transfer. With this approximation, 
and taking arbitrarily the atom chemical
potential to be zero on the facet, we get for the chemical potential
difference for an atom going from the tip to the facet :
\begin{equation}
\Delta \mu = \gamma_{surface} \kappa 
\label{Dmu} 
\end{equation}
where $\gamma_{surface}$ is the average surface tension on the tip and
$\kappa$ its curvature. Finally, we obtain the free energy barrier for
nucleation  :

\begin{equation}
\Delta G^* = \frac{\pi \gamma_{line}^2}{\Delta \mu}=\frac{\pi \gamma_{line}^2}{\gamma_{surface}} \frac{1}{\kappa}
\label{gk}
\end{equation}

\noindent
Figure \ref{dg1k} shows that Eqs. \ref{Dmu} and \ref{gk}
are in good agreement with our simulations and give coherent values for
$\gamma_{surface}$, close to the binding energy $E=0.1 eV$ (Eq. \ref{Dmu} 
gives $0.179 \pm 0.008$ eV atom$^{-2}$ and Eq. \ref{gk} leads to 
$0.175 \pm 0.035$ eV atom$^{-2}$).

\vspace{.4cm}
\fbox{\it Conclusion, Perspectives}
\vspace{.4cm}

The physical picture of nanocrystallite equilibration is the following
: above the roughening temperature, the continuous approach works well 
and leads to the classic fourth power law, the mass transfer being via 
atomic diffusion from
kinks or steps from the high curvature regions to the existing kinks
or steps of the low curvature region (which act as traps).  Below this
temperature however, large facets do appear in the low curvature
regions and no kinks or steps are available, preventing the diffusing
atoms from sticking there. Therefore, the route to equilibrium has to
involve nucleation of new atomic planes, which is much more difficult
and needs more time, leading to an exponential increase of $t_{nuc}$
(and therefore $t_{eq}$, which is directly related) as a function of
the cluster size. We actually see no reason why this picture would not
apply to much larger particles, up to the micrometer range. If
defect-free particles of this size could be produced, below the
roughening temperature they should show perfectly flat facets which
would demand the nucleation of a germ for effective mass transfer,
thus generating an exponential-type size dependence, and probably
preventing any experimental observation of the equilibration 
\cite{metois}! Many open questions remain : the temperature dependence of
$t_{eq}$ has to be understood, a more quantitative theory for $\Delta
G^*$ has to be worked out, and simulations in other geometries
(including also the substrate) would also be of interest.

We acknowledge useful discussions with H. Larralde, J.L. Barrat, L. 
Bocquet, J.J. M\'etois and J. Krug.

\begin{figure} 
\caption{Log-log dependence of the relaxation time as a function of the size
of the crystallites for different temperatures. The slope of each
linear fit is indicated.}
\label{teq_N}
\end{figure}

\begin{figure}
\caption{Morphologies of crystallites of 1728 atoms at two different 
temperatures : (a) partially rough at 700 K and (b)
fully faceted at 300 K. The colour of each atom depends on its number
of neighbors.}
\label{morpho}
\end{figure}

\begin{figure} 
\caption{Evolution of the total energy of crystallites as a 
function of the time logarithm for a 1728-atom cluster at two
temperatures. The arrows in the low temperature curve indicate the
transitions from one faceted configuration to the next. The total energy is
defined as the number of atomic bonds times the bond energy (E=0.1
eV). At the end of each curve, the crystallite has almost reached its
equilibrium shape.}
\label{energy}
\end{figure}

\begin{figure} 
 \caption{Cluster free energy during the formation of a nucleation
 germ on a facet as a function of the number $q$ of atoms in the
 germ. The curves have been obtained at 400 K, for several cluster sizes 
which have approximately the same shape, close to equilibrium (their aspect
 ratio is indicated in the figure). Clearly, the free energy barrier for 
the nucleation of a critical germ becomes larger as the crystallite size increases. Each solid curve is fitted by Eq. \ref{fit}, allowing to obtain
$\gamma_{line}$ and $\Delta \mu$.}
\label{deltag} 
\end{figure} 

\begin{figure} 
 \caption{Dependence of the nucleation barrier (given by the maximum
 of the curves in Fig. \protect\ref{deltag}) on the tip
 curvature $\kappa$ (calculated from the crystallite shape). In the
 inset, we show the dependence of $\Delta \mu$ (deduced from the
 fits in Fig. \protect\ref{deltag}) on the curvature $\kappa$. We fit 
these two curves by Eqs. \protect\ref{gk} and \protect\ref{Dmu} respectively,
obtaining $\gamma_{surface}$ (see text).}
\label{dg1k} 
\end{figure}


\begin{thebibliography}{99}

\bibitem{exp}
C. R. Stoldt et al., Phys. Rev.  Lett. {\bf 81}, 2950 (1995); 
A. Ichimiya, Y. Tanaka and K. Hayashi, Surf. Rev. and Lett. {\bf 5},
821 (1998).

\bibitem{review}
P. Jensen, Rev. Mod. Phys. {\bf 71}, 1695 (1999).

\bibitem{nichols} 
C. Herring, Phys Rev {\bf 82} 87 (1951); F.A. Nichols and W.W. Mullins, 
J. Appl. Phys., {\bf 36}, 1826 (1965); W.W. Mullins, Metall. and Mat.
Trans. A {\bf 26}, 1917 (1995).

\bibitem{expmull}
M. Drechsler et al. Journal de Physique {\bf 50}, Colloque {\bf C8},
223 (1989); H. P. Bonzel and E. E. Latta, Surf. Sci. {\bf 76}, 275 (1978);
G. Jeffers, M. A. Dubson and P. M. Duxbury, J. Appl. Phys. {\bf 75}, 5016 
(1994); R. Thouy, N. Olivi-Tran and R. Jullien, Phys. Rev B {\bf 56}, 5321 
(1997); J. Eggers, Phys Rev Lett. {\bf 80} 2634 (1998).

\bibitem{alb}
A.-L. Barab\'asi and H. E. Stanley, 1995, {\it Fractal Concepts in Surface
Growth} (Cambridge University Press); J. Lapujoulade Surf. Sci. Rep. {\bf 20}
191 (1994).

\bibitem{rough_cont}
H. Spohn, J. Phys. I France {\bf 3} 69 (1993).

\bibitem{kmc} 
A. B. Bortz, M. H. Kalos and J. L. Lebowitz, J. Comp. Phys.  {\bf 17}
10 (1975); A. F. Voter, Phys. Rev. B {\bf 34}, 6819 (1986).

\bibitem{last_one} 
H. Shao, P. C. Weakliem and H. Metiu, Phys. Rev. B {\bf 53}, 16041 (1996); 
P. Jensen et al. Eur. Phys. J. B {\bf 11}, 497 (1999).

\bibitem{note1}
To accelerate our simulations, only particles with less than 7
neighbors are allowed to move : this approximation is especially
justified at low temperatures, since motion of atoms with more than 6
neighbors become very rare (there are always some particles
with 6 neighbors which move much faster).

\bibitem{bogicevic}
A. Bogicevic, J. Str\"omquist and B. Lundqvist, Phys. Rev. Lett. 
{\bf81}, 637 (1998).

\bibitem{es}
R. L. Schwoebel, J. Appl. Phys. {\bf 40}, 614 (1969); R. L.  Schwoebel
and E. J. Shipsey, J. Appl. Phys. {\bf 37}, 3682 (1966).

\bibitem{note3}
The aspect ratio is defined as the maximum of all the possible ratios obtained
with the gyration radii along the $x$, $y$ and $z$ axis. Actually, it turns out 
that the initial aspect ratio is not
important since the relaxation rate dramatically slows down as the
crystallite approaches equilibrium (see Fig. \protect\ref{energy}),
and the equilibration time is dominated by the final steps.

\bibitem{nucleation}
See for example, {\it Solids Far from Equilibrium}, C. Godr\`eche (Ed.) 
Cambridge Univ Press (1992); A. Pimpinelli and J. Villain {\it Physics of 
Crystal Growth} (Cambridge University Press, 1998).

\bibitem{umbrella} 
D. Chandler {\it Introduction to Modern Satistical Mechanics} (Oxford
University Press, 1987).

\bibitem{note2} 
Unfortunately, the umbrella technique gives the
free energy to an additive constant. Moreover, the values obtained
for $q_0 < 2$ are unphysical since no germ exists. Therefore to be
able to compare our curves, we fit them in Fig. \protect\ref{deltag}
with Eq. \protect\ref{fit} (which works very well for $q_0 > 2$), and
we extrapolate the fit down to $q_0=0$ to find the free energy value
without germ. The additive constant is then fixed for each curve by
choosing $\Delta G (q=0, any \ N) \equiv 0$.

\bibitem{metois}
J.J. M\'etois and J.C. Heyraud, J. Crystal Growth {\bf 57}, 487 (1982). 

\end{thebibliography}
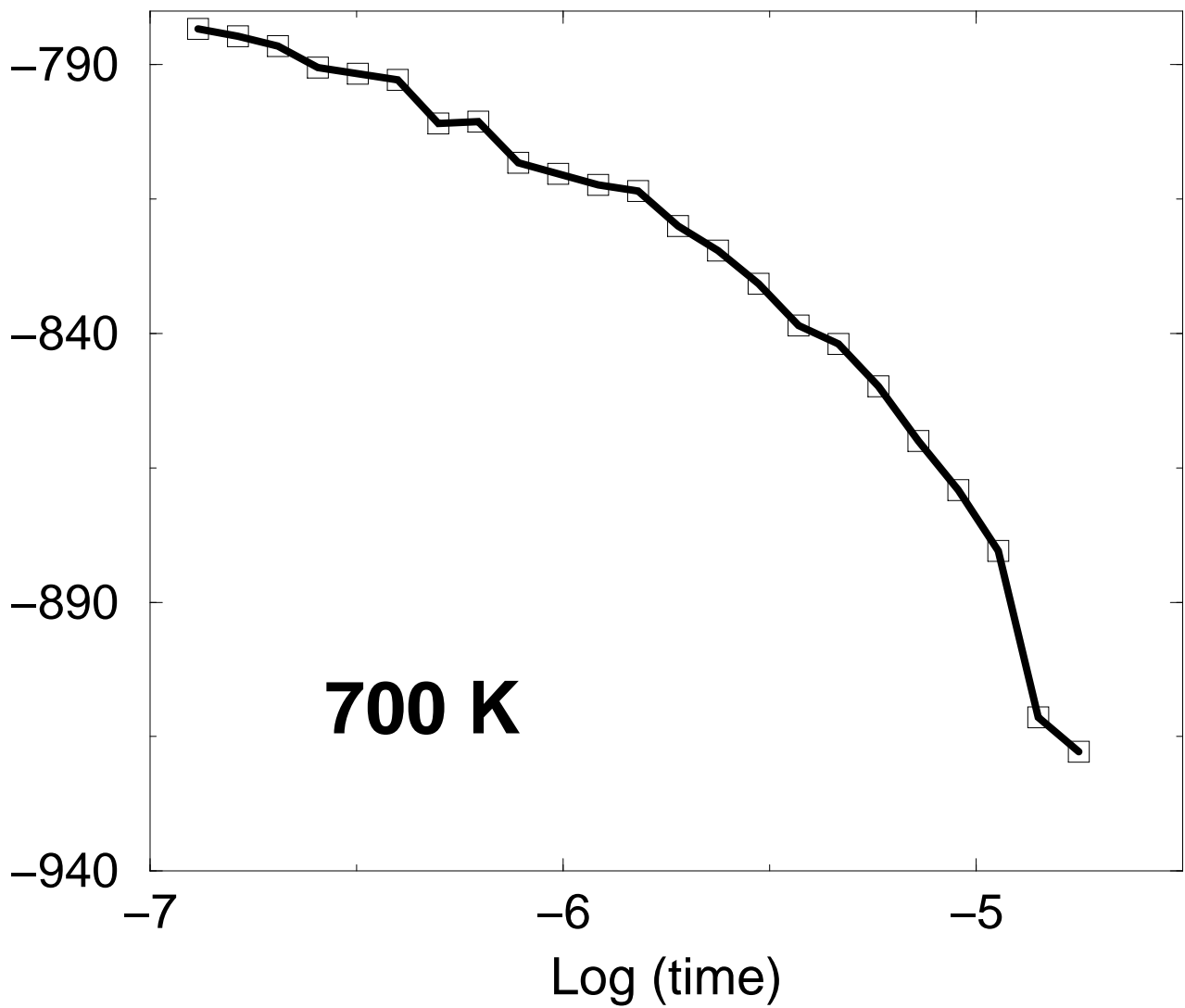
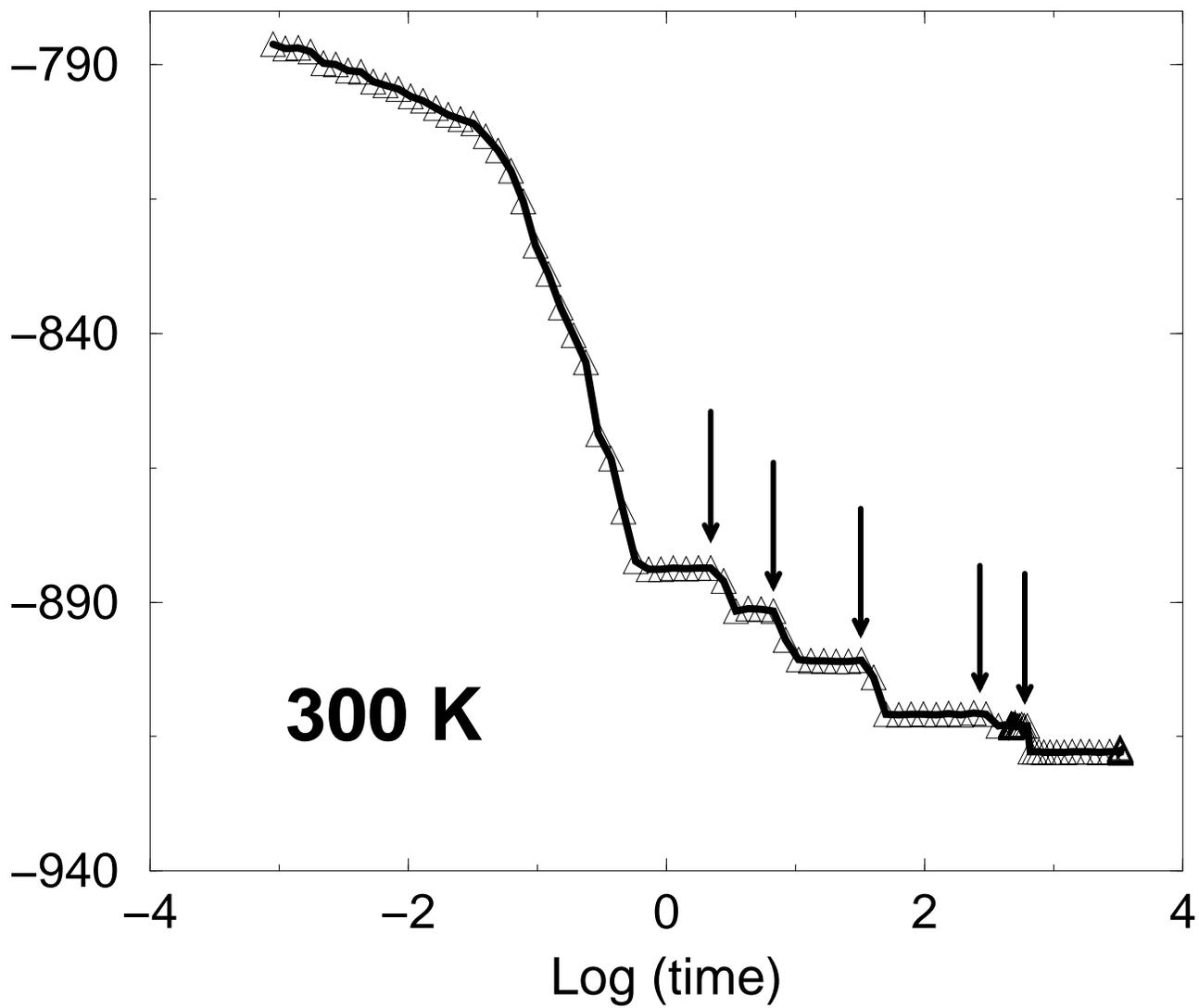
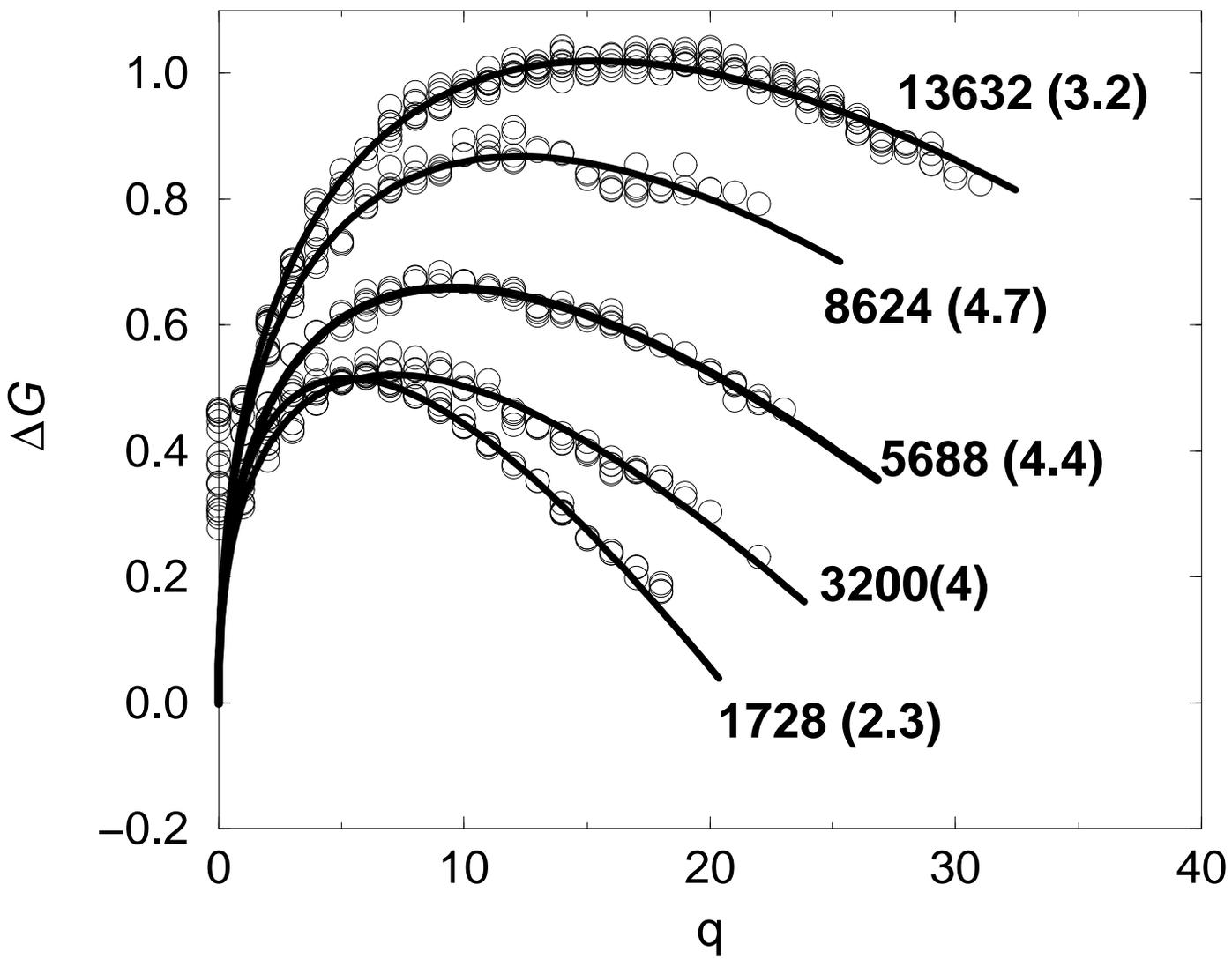
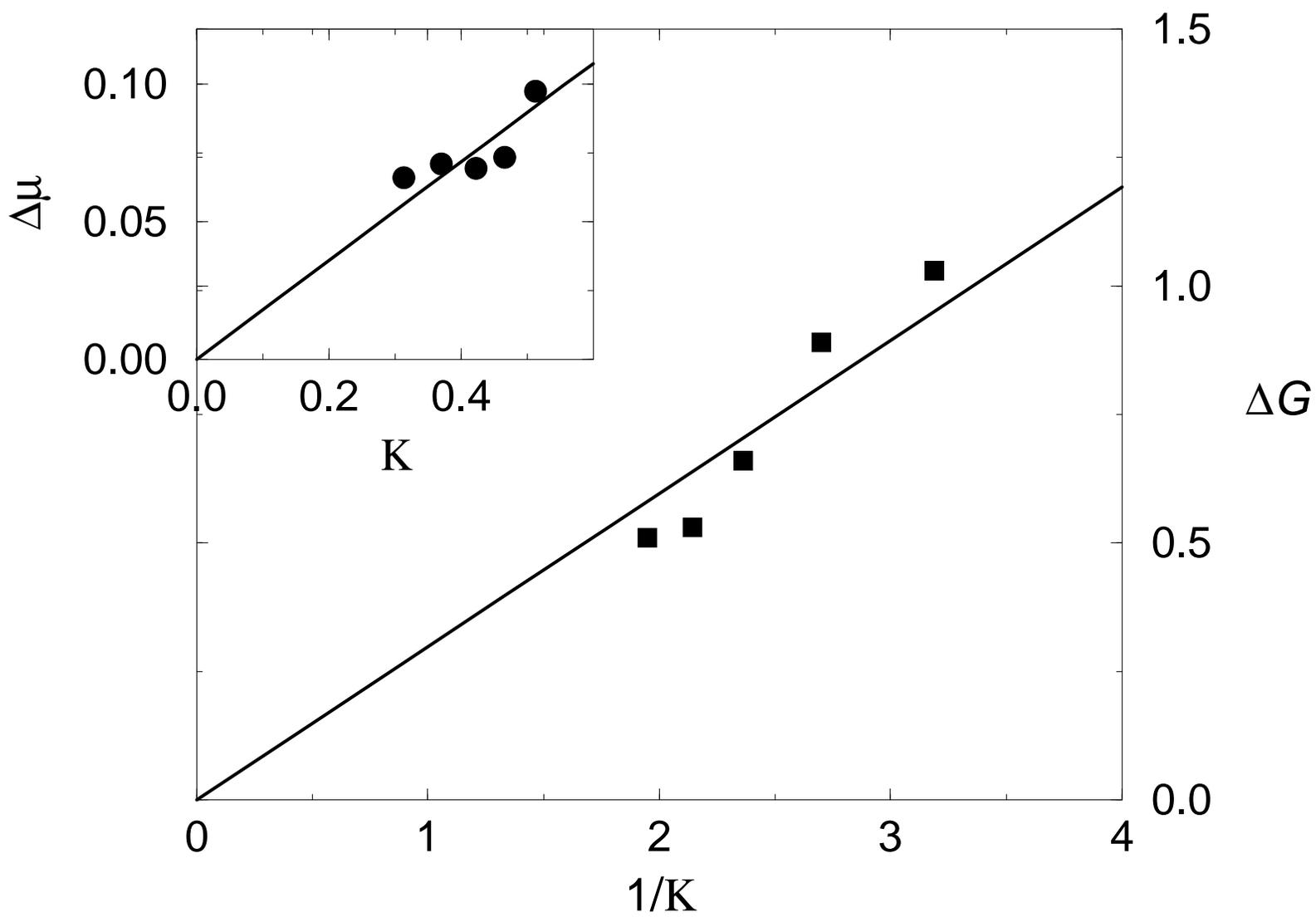
\end{document}